\documentclass{article}
\usepackage{setspace}
\usepackage{geometry}
\usepackage{amsmath}
\usepackage{amsthm}
\usepackage{amssymb}
\usepackage{lscape}
\usepackage{hyperref}
\usepackage[all,pdf]{xy}
\usepackage{indentfirst}
\usepackage{cite}
\usepackage{graphicx}
\newcommand{\lt}{<}
\newcommand{\gt}{>}
\hypersetup{hidelinks}
\newtheorem{theorem}{Theorem}
\newtheorem{remark}{Remark}

\newtheorem{condition}{Condition}

\newtheorem{assumption}{Assumption}
\DeclareMathOperator*{\argmax}{arg\,max}
\DeclareMathOperator*{\argmin}{arg\,min}
\title{Propensity Score Adapted Covariate Selection for Causal Inference}
\author{
 Kangjie Zhou\thanks{Department of Statistics, Stanford University} 
 \and 
 Jinzhu Jia\thanks{Department of Biostatistics, School of Public Health, Peking University}
}
\date{}

\begin{document}
	
\maketitle

\begin{abstract}
	In this paper, we propose a propensity score adapted variable selection procedure to select covariates for inclusion in propensity score models, in order to eliminate confounding bias and improve statistical efficiency in observational studies. Our variable selection approach is specially designed for causal inference, it only requires the propensity scores to be $\sqrt{n}$-consistently estimated through a parametric model and need not correct specification of potential outcome models. By using estimated propensity scores as inverse probability treatment weights in performing an adaptive lasso on the outcome, it successfully excludes instrumental variables, and includes confounders and outcome predictors. We show its oracle properties under the ``linear association" conditions. We also perform some numerical simulations to illustrate our propensity score adapted covariate selection procedure and evaluate its performance under model misspecification. Comparison to other covariate selection methods is made using artificial data as well, through which we find that it is more powerful in excluding instrumental variables and spurious covariates.
\end{abstract}

\section{Introduction}
\noindent When estimating the magnitude of causal effect of treatment assignment (exposure) on outcome variables in the presence of confounding factors from observational data, researchers usually rely on the unconfoundedness assumption (Rosenbaum and Rubin, 1983)\cite{RosenbaumRubin1983}, which states that there are no unmeasured confounders in covariates to be included in the model. Confounders are covariates that are associated with both exposure and outcome, and violation of the unconfoundedness assumption is often regarded as a serious fault since this can lead to inconsistent estimators of average treatment effect (ATE). To avoid such confounding bias, it is suggested to include all confounders into the regression model of treatment or outcome in real practice.

The propensity score method, first proposed in Rosenbaum and Rubin (1983)\cite{RosenbaumRubin1983}, has been widely used to construct consistent estimators of ATE. Propensity score, known as the conditional probability of receiving treatment given covariates, is usually estimated nonparametrically or semiparametrically to obtain asymptotically efficient estimators of ATE, even if the true propensity score is known. For example, the IPW estimator (Hirano et al., 2003)\cite{Hirano2003}. Some other efficient estimator also estimate the regression model of potential outcomes simultaneously, like the imputation estimator proposed in Hahn (1998)\cite{Hahn1998}. Here, we say an estimator of ATE is efficient, if and only if its asymptotic variance attains the semi-parametric efficiency bound (Hahn, 1998)\cite{Hahn1998}. It is important to determine which covariates should be included into propensity score model or regression models of potential outcomes, since this is closely related to asymptotic performance of estimators, there are two types of covariates that can influence statistical efficiency:
\begin{itemize}
	\item Instrumental variables: they are related to exposure but not outcome, inclusion of instrumental variables into the model will lead to variance inflation. See Hahn (2004)\cite{Hahn2004} for exclusion restriction in the outcome relation and Zhou et al. (2018) for an extended argument. 
	
	\item Outcome predictors: they are associated with outcome, but irrelevant to exposure, inclusion of outcome predictors into the model is at least innocuous, and helps improve statistical efficiency for some estimators. This is verified by simulations in Brookhart (2006)\cite{Brookhart2006} and proved in Zhou et al. (2018).
\end{itemize}

Therefore, an optimal set of covariates should include confounders to eliminate bias, and include outcome predictors and exclude instrumental variables, to reduce variance. Efficient variable selection methods should be proposed to achieve this object, and there has been a vast literature dedicated on discussing this problem, for instances, see Robins et al. (1986)\cite{robins1986role}, Vansteelandt et al. (2012)\cite{vansteelandt2012model} and Van der Laan et al. (2010)\cite{van2010collaborative}. De Luna et al. (2011)\cite{DeLuna2011} highlighted the significance of dimension reduction in nonparametric estimation of ATE and proposed an approach for identification of minimal sets of covariates using graphical models. Ertefaie et al. (2018)\cite{ertefaie2018variable} utilized a simultaneous penalization method to account for the relationship between weak confounders and outcome and treatment, which produced a lasso-type estimator whose oracle properties were shown. Shortreed et al. (2017)\cite{Shortreed2017} designed the outcome-adaptive lasso to perform both covariate selection and causal effect estimation, they chose tuning parameter through minimizing the wAMD for the sake of calculating the IPW estimator, which differed from the GCV method suggested by Tibshirani (1996)\cite{Tibshirani1996} and Fan et al. (2001)\cite{FanLi2001} and aligned more closely with causal inference. Besides, there are also a variety of Bayesian methods concerning on variable selection for causal inference, for instances, the Bayesian adjustment for confounding (BAC) introduced by Wang et al. (2012)\cite{wang2012bayesian}, and a decision-theoretic approach to confounder selection proposed by Wilson et al. (2014)\cite{wilson2014confounder}.

In this paper, we propose a propensity score adapted covariate selection (PACS) procedure that is specifically designed for causal inference. Compared to most existing methods that usually assume a linear model for the outcome, our approach is robust to outcome model misspecification. After successfully selecting the confounders and outcome predictors from the covariates, we use them to estimate propensity scores through a logistic model and then utilize an IPW estimator to estimate the average treatment effect. We also perform numerical simulations to show that our method is efficient in both covariate selection and ATE estimation.

The rest of this article is organized as follows. In section 2 we briefly review some previous results in variable selection and causal inference, then we will clarify the motivation of covariate selection methods in observational studies and the rationale behind our PACS. Then we introduce the PACS and the ``linear association" conditions formally in section 3. Its oracle properties are proven under mild regularity conditions, without assuming correct specification of outcome models. Details about implementing the algorithm will also be specified. We perform simulation studies in section 4 to demonstrate PACS's robustness to outcome model misspecification and compare its consistency in covariates selection and accuracy of ATE estimation with other existing methods. A concluding remark will be made in section 5.

\section{Variable Selection for Causal Inference}

\subsection{Causal Inference}
\noindent We use standard notation in causal inference, let $Y$ denote the outcome of a binary treatment $D$ in an observational study with $n$ individuals indexed by $i = 1, \cdots, n$. For each individual $i$, let $(Y_i^T, Y_i^C)$ denote the potential outcomes corresponding to treatment and control, respectively, then the observed outcome can be expressed as $Y_i = D_i Y_i^T + (1-D_i) Y_i^C$. The fundamental problem of causal inference is that we can never observe $Y_i^T$ and $Y_i^C$ simultaneously, but we aim to estimate the average treatment effect
\begin{equation}
	\text{ATE} = \mathrm{E} \lbrack Y^T \rbrack - \mathrm{E} \lbrack Y^C \rbrack,    \tag*{}
\end{equation}
sometimes under the presence of confounding factors. Hence, this can also be viewed as a missing data problem.

Let $\mathbf{X}$ denote the vector of covariates that helps in prediction of $Y$, then the propensity score (Rosenbaum and Rubin, 1983)\cite{RosenbaumRubin1983} is defined as the probability of receiving treatment given $\mathbf{X}$, i.e.,
\begin{equation}
	p(\mathbf{X}) = \mathrm{P}(D=1 \vert \mathbf{X}) = \mathrm{E}\lbrack D \vert \mathbf{X} \rbrack,    \tag*{}
\end{equation}
propensity score methods, like weighting, blocking and matching, have resulted in a large number of consistent estimators of ATE, see Imbens (2004)\cite{Imbens2004} for a complete review. When using propensity score methods, prudently selecting covariates to be included is important. Generally speaking, there are three types of covariates that need to be considered:
\begin{itemize}
	\item Treatment predictors (instrumental variables), which are related to exposure but not to outcome, unless through exposure. We use $\mathbf{I}$ to denote instrumental variables.
	\item Confounders, which are associated with both treatment and outcome, usually denoted by $\mathbf{U}$.
	\item Outcome predictors, which are pretreatment characteristics denoted by $\mathbf{C}$ and unaffected by $D$. They are correlated to outcome.
\end{itemize}
See the following directed acyclic graph (DAG) in Figure 1 for the three types of covariates aforementioned.
\begin{equation}
\xymatrix{
	& & \mathbf{U} \\
	\mathbf{I} & D & & Y & \mathbf{C}
	\ar"2,1";"2,2",
	\ar"2,2";"2,4",
	\ar"2,5";"2,4",
	\ar"1,3";"2,2",
	\ar"1,3";"2,4"
}                          \tag*{}
\end{equation}
\begin{center}
	Figure 1. The Directed Acyclic Graph for $(Y, D, \mathbf{I}, \mathbf{U}, \mathbf{C})$.
\end{center}
Denote $\mathbf{X}=(\mathbf{I}, \mathbf{U}, \mathbf{C})$, in here and sequel, we make the following assumptions for the instrumental variables:
\begin{assumption}
	The instrumental variables are independent of confounders and outcome predictors, i.e., $\mathbf{I} \perp (\mathbf{U}, \mathbf{C})$. This assumption is important in instrumental variable estimations.
\end{assumption}
\begin{assumption}
	(Exclusion Restriction) $Y^T \perp \mathbf{I}$, $Y^C \perp \mathbf{I}$. Some literature assumes $(Y^T, Y^C) \perp \mathbf{I}$, which is a stronger version of this assumption.
\end{assumption}
\begin{remark}
	In graphical causal models, an alternative of assumption 2 is specified as $Y \perp \mathbf{I} \vert D, \mathbf{U}, \mathbf{C}$, which can also be proved directly through the probability representation of figure 1. Assumption 2 is usually applied in potential outcome frameworks and is actually a formalization of the assumption that the counterfactual variable $Y(d, i)$ does not depend on $i$. Here, $Y(d, i)$ represents the outcome, which may be contrary to the fact, if the treatment is assigned to $D=d$ and the instrumental variables $\mathbf{I}=i$.
\end{remark}
In observational studies, the following ignorability of treatment assignment mechanism (Rosenbaum and Rubin, 1983)\cite{RosenbaumRubin1983} should be assumed to ensure that there are no unmeasured confounders and the treatment assignment is not deterministic.
\begin{assumption}
	The ignorability assumption of treatment assignment mechanism:
	\begin{itemize}
		\item[(i)] Unconfoundedness assumption: $(Y^T, Y^C) \perp D \vert \mathbf{X}$.
		\item[(ii)] Positivity assumption: $0 \lt p(\mathbf{X}) \lt 1$ for a.s. $\mathbf{X}$.
	\end{itemize}
\end{assumption}
\begin{remark}
	In fact, Rosenbaum and Rubin (1983)\cite{RosenbaumRubin1983} showed that $(Y^T, Y^C) \perp D \vert p(\mathbf{X})$ (unconfoundedness given the propensity score), and the propensity score function $p(\mathbf{X})$ is the coarsest balancing score. Here, a measurable function $b(\mathbf{X})$ of $\mathbf{X}$ is called a balancing score, if $(Y^T, Y^C) \perp D \vert b(\mathbf{X})$.
\end{remark}
According to assumption 3, to obtain consistency of estimators of average treatment effect one must include all confounders into the propensity score model. To achieve this goal, practitioners usually include all potential confounders, sometimes may contain instrumental variables, into the covariates to make sure that assumption 3 holds. However, recent studies have highlighted the variance inflation caused by inclusion of covariates that are related to treatment, but not outcome, e.g., see de Luna et al. (2011)\cite{DeLuna2011} and Patrick et al. (2011)\cite{Patrick2011}. It is also noteworthy that incorporating outcome predictors into estimation of propensity score helps improve statistical efficiency, although outcome predictors are in fact not in the propensity score model. For more details, please refer to Brookhart et al. (2006)\cite{Brookhart2006} for simulation evidences and Zhou et al. (2018) for a rigorous proof.

However, there is no difference between the three types of covariates in traditional formulation of the propensity score, and the significance of variable selection for causal inference has not received as enough attention as it should have been. Moreover, most of existing propensity score methods that are nonparametric or semiparametric and designed for estimation of ATE do not require the outcome model to be correctly specified, while in the same time some of them are still asymptotically efficient. Therefore it's nature to expect a consistent covariate selection approach possessing robustness to model misspecification as well. Motivated by these aforementioned facts and recent works in this field, we proposed the PACS which is an oracle procedure and robust to outcome model misspecification.

\subsection{Variable Selection}
The fundamental goal of variable selection is to identify a true model for the sake of accurate prediction of the outcome. In causal inference study, consider the outcome variable $Y$, and the vector of covariates $\mathbf{X}=(X_1, \cdots, X_p)$, where $p$ denotes the number of covariates in the entire model. We aim to identify the confounders and outcome predictors that helps eliminate bias and reduce variance, leading to efficient estimation of ATE using this selected model. Before introducing our PACS, we first review the adaptive lasso in Zou (2006)\cite{Zou2006}.

The adaptive lasso is an extension of the traditional lasso (Tibshirani, 1996)\cite{Tibshirani1996} which does not enjoy the oracle properties. It uses weights specified by $\sqrt{n}$-consistent estimators of the true parameter to force variables that are actually not in the model to be excluded. Under certain regularity conditions, Zou (2006)\cite{Zou2006} showed that the adaptive lasso satisfies the oracle properties, i.e., both consistency in covariate selection and asymptotic normality of the regularized estimator, if the model is correctly specified.

Consider an outcome $Y$ with $p$ predictors $\mathbf{X}=(X_1, \cdots, X_p)$, where the background may not necessarily be causal inference study. Let $\beta$ be the true parameter and $l_n(\beta; Y, \mathbf{X})$ be the log-likelihood function of a sample with size $n$. Set $\lambda_n \gt 0$ as a tuning parameter, then the adaptive lasso estimator is defined as the solution of the following optimization problem:
\begin{equation}
	\hat{\beta}_{AL} = \argmin_{\beta} \left\{ -l_n(\beta; Y, \mathbf{X}) + \lambda_n \sum_{j=1}^{p} \hat{\omega}_j \vert \beta_j \vert \right\}, \tag*{}
\end{equation} 
where $\hat{\omega}_j = \vert \tilde{\beta}_j \vert^{-\gamma}$ such that $\gamma \gt 0$, $\tilde{\beta}$ is a $\sqrt{n}$-consistent estimator of $\beta$, for example, the unpenalized maximum likelihood estimator. In most occasions, $l_n(\beta; Y, \mathbf{X}) = -\lVert Y-\mathbf{X} \beta \rVert^2$, hence $\tilde{\beta}$ is the OLS estimator.

The robustness of adaptive lasso to model misspecification is theoretically studied in Lu et al. (2012)\cite{lu2012robustness}. If the outcome model is misspecified, there may probably not exist a well-defined parameter $\beta$, instead of which the least false estimator $\beta^*$ is denoted as the quantity to which $\tilde{\beta}$ converges under regularity conditions. Similar results as oracle properties can be established if $\beta^*$ has some kind of sparse structure. Motivated by this observation, we meticulously design the form of $l_n(\beta; Y, \mathbf{X})$ such that the components of $\beta^*$ corresponding to instrumental variables and spurious covariates equal to $0$, and those corresponding to outcome predictors and confounders do not equal to $0$, in the context of observational studies. To achieve this, propensity scores should be $\sqrt{n}$-consistently estimated through a parametric model and inversed as the weights in a penalized linear regression of $Y$ on $\mathbf{X}$. Specifically, we propose the propensity score adapted covariate selection (PACS) procedure in next section.
\section{The Propensity Score Adapted Covariate Selection}
\noindent Following previous notation, let $\mathcal{U}$ denote indices of covariates that are associated with both outcome and exposure, i.e., the confounders, and $\mathcal{C}$ denote indices of outcome predictors that are unaffected by treatment assignment. Furthermore, we also utilize $\mathcal{I}$ to denote indices of treatment predictors that are unassociated with outcome, i.e., instrumental variables, and $\mathcal{S}$ denote spurious covariates uncorrelated to both outcome and exposure, i.e., not included in the model. The aim of propensity score model selection is to include $\mathbf{X}_{\mathcal{U}}$ and $\mathbf{X}_{\mathcal{C}}$, and exclude $\mathbf{X}_{\mathcal{I}}$ and $\mathbf{X}_{\mathcal{S}}$ simultaneously. To achieve this, we propose a propensity score adapted covariate selection (PACS) procedure that puts an adaptive lasso penalty on weighted linear regression of $Y$ on $\mathbf{X}$. The OLS weights are specified by propensity scores, which need to be $\sqrt{n}$-consistently estimated through a parametric model at the beginning.
\begin{remark}
	Note that although we did not mention $\mathbf{X}_{\mathcal{S}}$ in previous section introducing causal inference, they appear quite frequently in real practice when there are a large amount of covariates, and a large proportion of which is irrelevant or weakly relevant to both treatment and outcome. Hence, it is necessary to design a variable selection method that is powerful in excluding spurious covariates. Since $\mathbf{X}_{\mathcal{S}}$ are not in the model, it is nature to make the assumption that $\mathbf{X}_{\mathcal{S}}$ is independent of the potential outcomes $(Y^T, Y^C)$ and the exposure $D$. Furthermore, we assume that $\mathbf{X}_{\mathcal{S}} \perp (\mathbf{X}_{\mathcal{U}}, \mathbf{X}_{\mathcal{C}})$. In coincidence with notation in section 2.1, we have $\mathbf{X}_{\mathcal{I}} = \mathbf{I}$, $\mathbf{X}_{\mathcal{U}} = \mathbf{U}$ and $\mathbf{X}_{\mathcal{C}} = \mathbf{C}$, for the covariates that are actually in the model.
\end{remark}

\subsection{Step 1: Estimate the Propensity Scores}
\noindent We first assume that there exists an estimator $\hat{p}(\mathbf{X})$ of the propensity score function $p(\mathbf{X})$, such that $\hat{p}(\mathbf{X})-p(\mathbf{X})=O_p(1/\sqrt{n})$. This is always true if a generalized linear model of propensity score is correctly fitted. In most common case, we assume a logistic model for the propensity score:
\begin{equation}
	\log \left( \frac{p(\mathbf{X})}{1-p(\mathbf{X})} \right) = \sum_{j \in \mathcal{I}} \alpha_j X_j + \sum_{j \in \mathcal{U}} \alpha_j X_j.  \tag*{}
\end{equation}
Then we estimate $\alpha$ through maximizing the following log-likelihood function:
\begin{equation}
	\hat{\alpha} = \argmax_{\alpha} \left\{ D_i \left(\mathbf{X}_i^\top \alpha\right) - \log\left(1+\exp\left(\mathbf{X}_i^\top \alpha\right)\right) \right\},  \tag*{}
\end{equation}
here $D_i$ is the treatment assignment of subject $i$, then $\hat{p}(\mathbf{X}_i) = \exp(\mathbf{X}_i^\top \hat{\alpha})/(1+\exp(\mathbf{X}_i^\top \hat{\alpha}))$ is a $\sqrt{n}$-consistent estimator of $p(\mathbf{X}_i)$ for all $i = 1, \cdots, n$.

\subsection{Step 2: Penalized Weighted Least-Squares Regression}
\noindent In this step, we do not assume that the outcome model, i.e., the relationship between $Y$ and $(D, \mathbf{X})$, is correctly specified. Consider the following weighted linear regression of $Y$ on $\mathbf{X}$ within the treatment group:
\begin{equation}
	\left(\tilde{\beta}^T, \tilde{\eta}^T\right) = \argmin_{\beta, \eta} \sum_{i \in T} \frac{1}{\hat{p}(\mathbf{X}_i)} \left( Y_i - \eta - \beta^\top \mathbf{X}_i \right)^2,  \tag{1}
\end{equation}
where $T$ denote the treatment group, $\hat{p}(\mathbf{X})$ is obtained from Step 1. Let $\lambda_n \gt 0$ be the tuning parameter, then we specify the weights as $\hat{\omega}_j^T = \vert \tilde{\beta}_j^T \vert^{-\gamma}$ for $\gamma \gt 0$, and define
\begin{equation}
    (\hat{\beta}_{PACS}^T, \hat{\eta}_{PACS}^T) = \argmin_{\beta, \eta} \sum_{i \in T} \frac{1}{\hat{p}(\mathbf{X}_i)} \left(Y_i - \eta - \beta^\top \mathbf{X}_i\right)^2 + \lambda_n \sum_{j=1}^{p} \hat{\omega}_j^T \vert \beta_j \vert,   \tag*{}
\end{equation}
Similarly, we may define $\tilde{\beta}^C$ for the control group, which is denoted by $C$:
\begin{equation}
	\left(\tilde{\beta}^C, \tilde{\eta}^C\right) = \argmin_{\beta, \eta} \sum_{i \in C} \frac{1}{1-\hat{p}(\mathbf{X}_i)} \left( Y_i - \eta - \beta^\top \mathbf{X}_i \right)^2,  \tag{2}
\end{equation}
and the weights $\hat{\omega}_j^C = \vert \tilde{\beta}_j^C \vert^{-\gamma}$ for $\gamma \gt 0$, and the corresponding tuning parameter $\lambda_n \gt 0$, then we have
\begin{equation}
	(\hat{\beta}_{PACS}^C, \hat{\eta}_{PACS}^C) = \argmin_{\beta, \eta} \sum_{i \in C} \frac{1}{1-\hat{p}(\mathbf{X}_i)} \left(Y_i - \eta - \beta^\top \mathbf{X}_i\right)^2 + \lambda_n \sum_{j=1}^{p} \hat{\omega}_j^C \vert \beta_j \vert.   \tag*{}
\end{equation}
For $j=1, \cdots, p$, if $\hat{\beta}_{PACS, j}^T \hat{\beta}_{PACS, j}^C \neq 0$, then we select covariate $X_j$ into the propensity score model.

\subsection{Oracle Properties of PACS}
\noindent Since $\hat{p}(\mathbf{X})$ is a $\sqrt{n}$-consistent estimator of $p(\mathbf{X})$, we know that both $\tilde{\beta}^T$ and $\tilde{\beta}^C$ converge to their corresponding least false parameters, $\beta^{T*}$ and $\beta^{C*}$, when the linear outcome model is unknown or misspecified. The oracle properties of $\hat{\beta}_{PACS}^T$ and $\hat{\beta}_{PACS}^C$ relied heavily on the sparsity of $\beta^{T*}$ and $\beta^{C*}$. Before presentation of our main result, we first introduce the ``linear association" conditions.

Denote $\mathcal{A} = \mathcal{U} \cup \mathcal{C}$, the indices of covariates that should be included in the model, and $\mathcal{A}^c = \mathcal{I} \cup \mathcal{S}$. Assume that $\vert \mathcal{A} \vert = p_0 \lt p$. For any two random vectors $\mathbf{U}$ and $\mathbf{V}$, let $\mathrm{Cov}(\mathbf{U}, \mathbf{V}) = \mathrm{E} \lbrack (\mathbf{U} - \mathrm{E}\lbrack \mathbf{U} \rbrack) (\mathbf{V} - \mathrm{E}\lbrack \mathbf{V} \rbrack)^\top\rbrack$ denote their cross-variance matrix. Additionally, we assume that $\mathrm{Cov} (\mathbf{X}_{\mathcal{A}}, \mathbf{X}_{\mathcal{A}})$ is an invertible $p_0 \times p_0$ matrix.
\begin{condition}[Linear Association Condition for Potential Outcome $Y^T$]
	For all $j=1, \cdots, p_0$,
	\begin{equation}
		\left(\mathrm{Cov} (\mathbf{X}_{\mathcal{A}}, \mathbf{X}_{\mathcal{A}})^{-1} \mathrm{Cov} (\mathbf{X}_{\mathcal{A}}, Y^T)\right)_j \neq 0.   \tag*{}
	\end{equation}
\end{condition}
\begin{condition}[Linear Association Condition for Potential Outcome $Y^C$]
	For all $j=1, \cdots, p_0$,
	\begin{equation}
	    \left(\mathrm{Cov} (\mathbf{X}_{\mathcal{A}}, \mathbf{X}_{\mathcal{A}})^{-1} \mathrm{Cov} (\mathbf{X}_{\mathcal{A}}, Y^C)\right)_j \neq 0.   \tag*{}
	\end{equation}
\end{condition}
\begin{remark}
	If the regression model of $Y^T$ is correctly specified as linear, i.e., there exists $\beta^T$ and $\eta^T$ such that
	\begin{equation}
		\mathrm{E} \left[ Y^T \vert \mathbf{X}_{\mathcal{A}} \right] = \eta^T + \mathbf{X}_{\mathcal{A}}^\top \beta^T,    \tag*{}
	\end{equation}
	then Condition 1 holds true. Similarly, if the regression model of $Y^C$ is correctly specified as linear, i.e., there exists $\beta^C$ and $\eta^C$,
	\begin{equation}
		 \mathrm{E} \left[ Y^C \vert \mathbf{X}_{\mathcal{A}} \right] = \eta^C + \mathbf{X}_{\mathcal{A}}^\top \beta^C,  \tag*{}
	\end{equation}
	then Condition 2 holds true. In fact, Condition 1 and Condition 2 are much easier to be satisfied than linear outcome models, because they only require the regression coefficients to be not equal to $0$.
\end{remark}
Now we can establish oracle properties of the PACS as an adaptive-lasso-type selector:
\begin{theorem}
	Suppose $\lambda_n/\sqrt{n} \rightarrow 0$ and $\lambda_n n^{(\gamma-1)/2} \rightarrow \infty$, for $\gamma \gt 0$, then under mild regularity conditions, if linear association condition for potential outcome $Y^T$ (Condition 1) holds, then
	\begin{itemize}
		\item[1.] $\lim_{n \rightarrow \infty} \mathrm{P} (\hat{\beta}_{PACS, j}^T \neq 0, \ \forall j \in \mathcal{A} = \mathcal{U} \cup \mathcal{C}) = 1$.
		\item[2.] $\lim_{n \rightarrow \infty} \mathrm{P} (\hat{\beta}_{PACS, j}^T=0, \ \forall j \in \mathcal{A}^c = \mathcal{I} \cup \mathcal{S}) = 1$.
		\item[3.] The limiting distribution of $\sqrt{n}(\hat{\beta}_{PACS}^T - \beta^{T*})$ is normal.
	\end{itemize}
    If linear association condition for potential outcome $Y^C$ (Condition 2) holds, then
	\begin{itemize}
		\item[1.] $\lim_{n \rightarrow \infty} \mathrm{P} (\hat{\beta}_{PACS, j}^C \neq 0, \ \forall j \in \mathcal{A} = \mathcal{U} \cup \mathcal{C}) = 1$.
		\item[2.] $\lim_{n \rightarrow \infty} \mathrm{P} (\hat{\beta}_{PACS, j}^C=0, \ \forall j \in \mathcal{A}^c = \mathcal{I} \cup \mathcal{S}) = 1$.
		\item[3.] The limiting distribution of $\sqrt{n}(\hat{\beta}_{PACS}^C - \beta^{C*})$ is normal.
	\end{itemize}
	Additionally, if both Condition 1 and Condition 2 hold, then we have
	\begin{itemize}
		\item[1.] $\lim_{n \rightarrow \infty} \mathrm{P} (\hat{\beta}_{PACS, j}^T \hat{\beta}_{PACS, j}^C\neq0, \ \forall j \in \mathcal{A} = \mathcal{U} \cup \mathcal{C}) = 1.$
		\item[2.] $\lim_{n \rightarrow \infty} \mathrm{P} (\hat{\beta}_{PACS, j}^T \hat{\beta}_{PACS, j}^C=0, \ \forall j \in \mathcal{A}^c = \mathcal{I} \cup \mathcal{S}) = 1.$
	\end{itemize}
\end{theorem}
(Proof in Appendix)
\begin{remark}
	If the ``linear association" conditions hold for both potential outcomes, $Y^T$ and $Y^C$, then our PACS is consistent in including confounders and outcome predictors, and more efficient in excluding instrumental variables and spurious covariates. Furthermore, the specific expression of variance of limiting distribution of $\sqrt{n}(\hat{\beta}_{PACS}^T - \beta^{T*})$ depends on the parametric model we apply to estimate propensity scores, this is also true for the limiting distribution of $\sqrt{n}(\hat{\beta}_{PACS}^C - \beta^{C*})$.
\end{remark}
\subsection{Parameter Selection and ATE Estimation}
\noindent We now discuss the computation issues of the PACS. In fact, the PACS can be transformed into an adaptive lasso problem, after centralization and reweighing on potential outcome variables and covariates. Hence, similar as Zou (2006)\cite{Zou2006}, we may use two-dimensional cross-validation to find an optimal pair of $(\lambda_n, \gamma)$ satisfying requirements $\lambda_n/\sqrt{n} \rightarrow 0$ and $\lambda_n n^{(\gamma-1)/2} \rightarrow \infty$, which is just the same as the adaptive lasso requires. Specifically speaking, if we define the following weighted averages of $Y$ and $\mathbf{X}$:
\begin{equation}
	\overline{Y}_p = \frac{\sum_{i=1}^{n} D_i Y_i / \hat{p}(\mathbf{X}_i)}{\sum_{i=1}^{n} D_i / \hat{p}(\mathbf{X}_i)}, \ \overline{Y}_{1-p} = \frac{\sum_{i=1}^{n} (1-D_i) Y_i / (1-\hat{p}(\mathbf{X}_i))}{\sum_{i=1}^{n} (1-D_i) / (1-\hat{p}(\mathbf{X}_i))},  \tag*{}
\end{equation}
and
\begin{equation}
    \overline{\mathbf{X}}_p = \frac{\sum_{i=1}^{n} D_i \mathbf{X}_i / \hat{p}(\mathbf{X}_i)}{\sum_{i=1}^{n} D_i / \hat{p}(\mathbf{X}_i)}, \ \overline{\mathbf{X}}_{1-p} = \frac{\sum_{i=1}^{n} (1-D_i) \mathbf{X}_i / (1-\hat{p}(\mathbf{X}_i))}{\sum_{i=1}^{n} (1-D_i) / (1-\hat{p}(\mathbf{X}_i))},  \tag*{}
\end{equation}
then $(1)$ is equivalent to
\begin{align*}
	\tilde{\beta}^T & = \argmin_{\beta} \sum_{i \in T} \frac{1}{\hat{p}(\mathbf{X}_i)} \left( Y_i-\overline{Y}_p - \beta^\top (\mathbf{X}_i - \overline{\mathbf{X}}_p) \right)^2  \\
	& = \argmin_{\beta} \sum_{i \in T} \left( \frac{Y_i-\overline{Y}_p}{\sqrt{\hat{p}(\mathbf{X}_i)}} - \beta^\top \frac{(\mathbf{X}_i - \overline{\mathbf{X}}_p)}{\sqrt{\hat{p}(\mathbf{X}_i)}} \right)^2.
\end{align*}
Therefore, in order to obtain $\hat{\beta}_{PACS}^T$, we only need to run an adaptive lasso of $(Y_i-\overline{Y}_p) / \sqrt{\hat{p}(\mathbf{X}_i)}$ on $(\mathbf{X}_i-\overline{\mathbf{X}}_p) / \sqrt{\hat{p}(\mathbf{X}_i)}$ in the treatment group $T$. We may apply Algorithm 1 (The LARS algorithm for the adaptive lasso) in Section 3.5 of Zou (2006)\cite{Zou2006}. Through this way, $\hat{\beta}_{PACS}^C$ can be efficiently computed as well. To perform numerical simulations, we implement function $\textsf{adalasso}$ in $\textsf{R}$ package $\textsf{parcor}$.

After the covariates need to be included into the propensity score model, denoted by $\mathbf{X}_{PACS}$, has been selected, we use them to calculate the Inverse Probability Weighting (IPW) estimator, which is of the form
\begin{equation}
	\widehat{\text{ATE}}_{ipw} = \frac{\sum_{i=1}^{n} D_i Y_i / \hat{p}(\mathbf{X}_{PACS, i})}{\sum_{i=1}^{n} D_i / \hat{p}(\mathbf{X}_{PACS, i})} -  \frac{\sum_{i=1}^{n} (1-D_i) Y_i / (1-\hat{p}(\mathbf{X}_{PACS, i}))}{\sum_{i=1}^{n} (1-D_i) / (1-\hat{p}(\mathbf{X}_{PACS, i}))},    \tag*{}
\end{equation}
whose statistical efficiency will be examined in next selection and regarded as an important criterion for evaluation of the propensity score adapted covariate selection procedure.

\section{Numerical Simulations}
\noindent In this section, we perform some numerical simulations with a main focus on model selection consistency of the PACS, since it has been discussed elaborately in Brookhart et al. (2006)\cite{Brookhart2006} that once the covariates beneficial for prediction accuracy of ATE, i.e., target covariates in Shortreed et al. (2017)\cite{Shortreed2017}, have been correctly selected, the resulting IPW estimator is more efficient than those using only confounders or all potential confounders, or only confounders and instrumental variables. We make a comparison between the PACS and the Outcome-adaptive lasso (abbreviated OAL) of Shortreed et al. (2017)\cite{Shortreed2017}, while the latter one was shown to outperform other variable selection approaches for causal inference, such as those mentioned in Section 1. We mainly consider the following two scenarios:
\begin{itemize}
	\item [1.] If the outcome model is misspecified, i.e., the regression model of $Y$ is not linear in $(D, \mathbf{X})$, our PACS is still consistent in covariate selection but the OAL is not. In this scenario, we assume that
	\begin{equation}
		Y_i^T = \mathbf{X}_i^\top \beta^T + \epsilon_i, \ Y_i^C = \mathbf{X}_i^\top \beta^C + \epsilon_i, \ \epsilon_i \sim \text{i.i.d.} \ \mathcal{N}(0, 1),   \tag*{}
	\end{equation}
	and the observed outcome is generated by the relationship $Y_i=D_i Y_i^T+(1-D_i)Y_i^C$ for $i=1, \cdots, n$.
	\item [2.] If the linear outcome model is correct, then our PACS performs better in excluding instrumental variables and spurious covariates than the OAL. In this scenario, we assume that
	\begin{equation}
		Y_i = \mathbf{X}_i^\top \beta + D_i \mu + \epsilon_i, \ \epsilon_i \sim \text{i.i.d.} \ \mathcal{N}(0, 1),    \tag*{}
	\end{equation}
	for $i=1, \cdots, n$, here $\mu$ is the magnitude of average treatment effect.
\end{itemize}
In both scenarios, we assume a logistic model for the propensity score, i.e., $p(\mathbf{X}) = \exp(\mathbf{X}^\top \alpha)/(1+\exp(\mathbf{X}^\top \alpha))$. Same as before, let $p$ denote the number of covariates, where $(X_1, \cdots, X_8)$ are actually in the model hence there are $p-8$ spurious covariates. Throughout our simulation, the first two covariates, $(X_1, X_2)$ are set to be confounders, while $(X_3, X_4)$ are outcome predictors, which means that they only appear in potential outcome models. The last four true covariates, $(X_5, X_6, X_7, X_8)$ are designated as instrumental variables that are only associated with exposure. We replicate $m$ datasets, each one of which have $n$ i.i.d. observed samples $(Y_i, D_i, \mathbf{X}_i), i=1, \cdots, n$. As for the distribution of $\mathbf{X}$, we assume that $\mathbf{X} \sim \mathcal{N}(0, I_p)$ for simplicity.

\subsection{Choice of Parameters}
\noindent We fix $m=200$ for the sake of computational convenience, and $(n, p) = (500, 20)$ in Scenario 1. In Scenario 2, we consider three different parameter combinations for $(n, p)$: $(n, p)=(500, 100)$ represents the case in which the ratio of number of covariates to sample size is large, $(n, p)=(500, 20)$ represents the case in which such ratio is small, and $(n, p)=(1000, 20)$ represents the case with large sample size. Note that the ratio $p/n$ can not be too large since the PACS requires to perform a logistic regression at the first stage.

In both scenarios, we consider the following two choices of $\alpha$ characterizing relationship between $D$ and $\mathbf{X}$:
\begin{itemize}
	\item [1.] $\alpha=(0.4, 0.4, 0, 0, 1, 1, 1, 1, 0, \cdots, 0)^\top$: in this choice we assume a weaker relationship between confounders and treatment assignment.
	\item [2.] $\alpha=(1, 1, 0, 0, 1, 1, 1, 1, 0, \cdots, 0)^\top$: in this choice we assume a stronger relationship between confounders and treatment assignment.
\end{itemize}
In scenario 2, $\mu$ and $\beta$ are set to be constant with varying sample size and number of covariates. To be concrete, we take $\mu=0$, $\beta=(0.6, 0.6, 0.6, 0.6, 0, \cdots, 0)^\top$. In scenario 1, we choose different combinations of $(\beta^T, \beta^C)$ to reflect various degrees of heterogeneity between the treatment group and the control group. Specifically, we consider the following choices respectively:
\begin{itemize}
	\item [1.] Small heterogeneity: $\beta^T=(0.6, 0.6, 0.8, 0.8, 0, \cdots, 0)^\top$, $\beta^C=(0.8, 0.8, 0.6, 0.6, 0, \cdots, 0)^\top$.
	\item [2.] Moderate heterogeneity: $\beta^T=(0.6, 0.6, 1.2, 1.2, 0, \cdots, 0)^\top$, $\beta^C=(1.2, 1.2, 0.6, 0.6, 0, \cdots, 0)^\top$.
	\item [3.] Great heterogeneity: $\beta^T=(0.6, 0.6, 2.4, 2.4, 0, \cdots, 0)^\top$, $\beta^C=(2.4, 2.4, 0.6, 0.6, 0, \cdots, 0)^\top$.
\end{itemize}
In next section, we present results of simulation studies to illustrate our PACS's robustness to outcome model specification, especially when there is a large difference between $\beta^T$ and $\beta^C$. We also demonstrate that PACS outperforms OAL in excluding instrumental variables and spurious covariates, when the linear outcome model is correct.

\subsection{Comparison of PACS and OAL}
\noindent We run $m=200$ recurrent simulations, then calculate and plot the frequency of each covariate being selected into the model for PACS and OAL, respectively. See Figure 2-5 for more details. In Scenario 1 ($\beta^T \neq \beta^C$), the PACS shows much lower error rate in excluding instrumental variables and spurious covariates, compared to the OAL. Specially, OAL selects instrumental variables and spurious covariates to the model in approximately $30\%-40\%$ of the $m=200$ attempts, when $\beta^T=(0.6, 0.6, 2.4, 2.4, 0, \cdots, 0)^\top$ and $\beta^C=(2.4, 2.4, 0.6, 0.6, 0, \cdots, 0)^\top$. In the meantime, PACS makes nearly no faults. Aside from its robustness to outcome model misspecification, PACS also outperforms OAL when the linear model is right, especially when there are a large number of covariates, e.g., $n=500$, $p=100$ as depicted in Figure 4 and Figure 5. As $n$ increases, PACS shows great stability in both cases whether the confounders is strongly or weakly linked to exposure.

We also argue that, PACS possesses better computational efficiency than OAL, when there is a large number of covariates or sample points. To implement the OAL algorithm, we use $\textsf{R}$ code $\textsf{OAL}$ in Supplementary Materials of Shortreed et al. (2017)\cite{Shortreed2017}. Consider the case $\alpha=(1, 1, 0, 0, 1, 1, 1, 1, 0, \cdots, 0)^\top$ in Scenario 2, when $n=500$, $p=100$ (many covariates), it takes OAL $308.60$ seconds to complete $m=200$ runs, while PACS only spends $26.69$ seconds. When $n=1000$, $p=20$ (large sample size), OAL needs $91.23$ seconds to finish $m=200$ cycles, while the running time of PACS is just $10.24$ seconds. See the following tables for more details:

\begin{center}
	\begin{tabular}{|c|c|c|c|}
		\hline
		$m=200$ & $n=500$, $p=100$ & $n=500$, $p=20$ & $n=1000$, $p=20$ \\
		\hline
		$PACS$ & $28.08s$ & $8.29s$ & $8.76s$ \\
		\hline
		$OAL$ & $396.02s$ & $54.50s$ & $84.74s$ \\
		\hline
	\end{tabular}
\end{center}

\begin{center}
	Table 1. Runtime Comparison: $\alpha=(0.4, 0.4, 0, 0, 1, 1, 1, 1, 0, \cdots, 0)^\top$ (weak confounders).
\end{center}

\begin{center}
	\begin{tabular}{|c|c|c|c|}
		\hline
		$m=200$ & $n=500$, $p=100$ & $n=500$, $p=20$ & $n=1000$, $p=20$ \\
		\hline
		$PACS$ & $26.69s$ & $6.61s$ & $10.24s$ \\
		\hline
		$OAL$ & $308.60s$ & $57.64s$ & $91.23s$ \\
		\hline
	\end{tabular}
\end{center}

\begin{center}
	Table 2. Runtime Comparison: $\alpha=(1, 1, 0, 0, 1, 1, 1, 1, 0, \cdots, 0)^\top$ (strong confounders).
\end{center}

\begin{landscape}
	\begin{center}
		$\beta^T=(0.6, 0.6, 0.8, 0.8, 0, \cdots, 0)^\top$ \hspace{26mm} $\beta^T=(0.6, 0.6, 1.2, 1.2, 0, \cdots, 0)^\top$ \hspace{26mm} $\beta^T=(0.6, 0.6, 2.4, 2.4, 0, \cdots, 0)^\top$ \\
		$\beta^C=(0.8, 0.8, 0.6, 0.6, 0, \cdots, 0)^\top$ \hspace{26mm} $\beta^C=(1.2, 1.2, 0.6, 0.6, 0, \cdots, 0)^\top$ \hspace{26mm} $\beta^C=(2.4, 2.4, 0.6, 0.6, 0, \cdots, 0)^\top$
	\end{center}
	\begin{center}
		\includegraphics[width=22em]{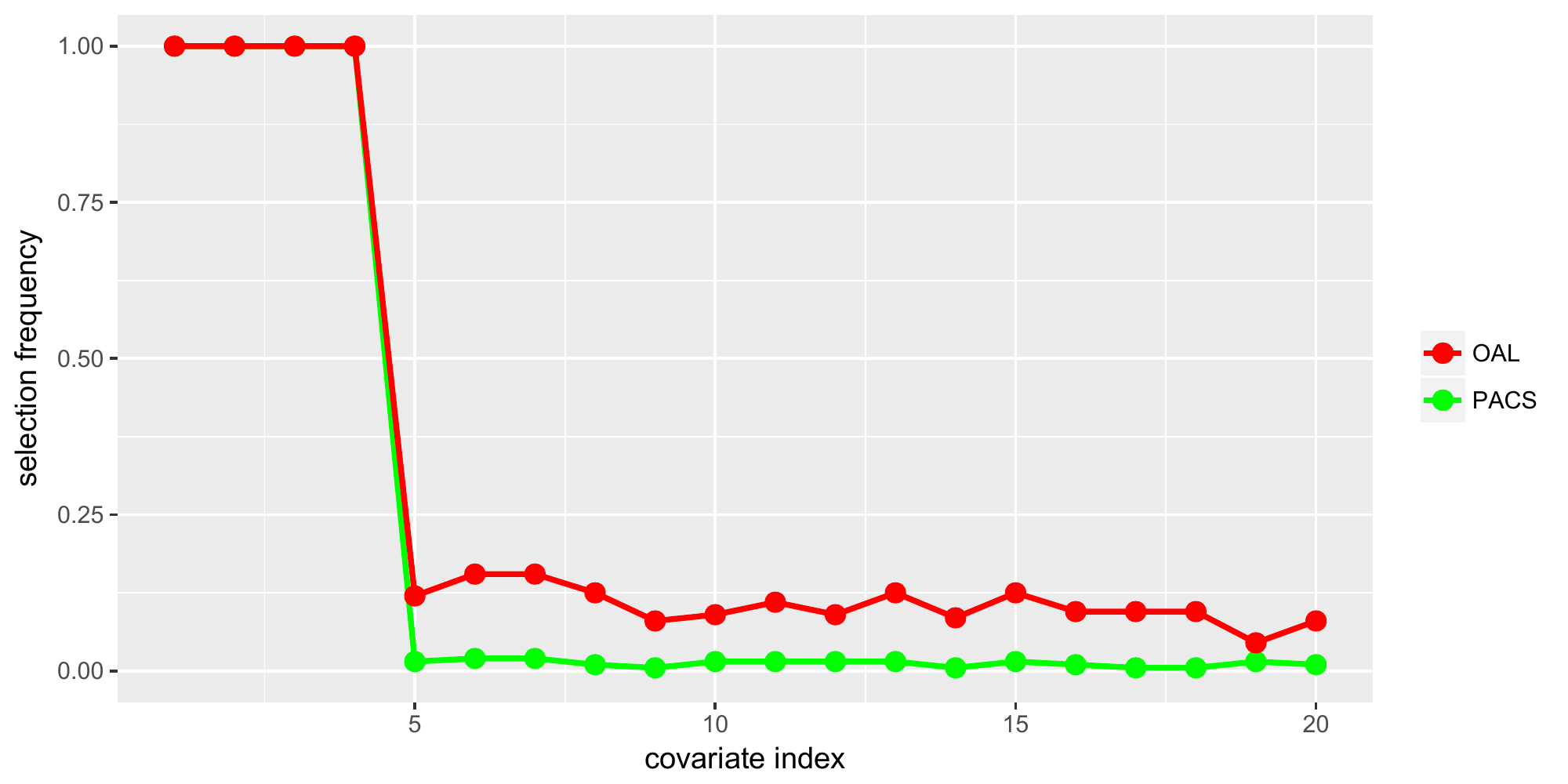}
		\includegraphics[width=22em]{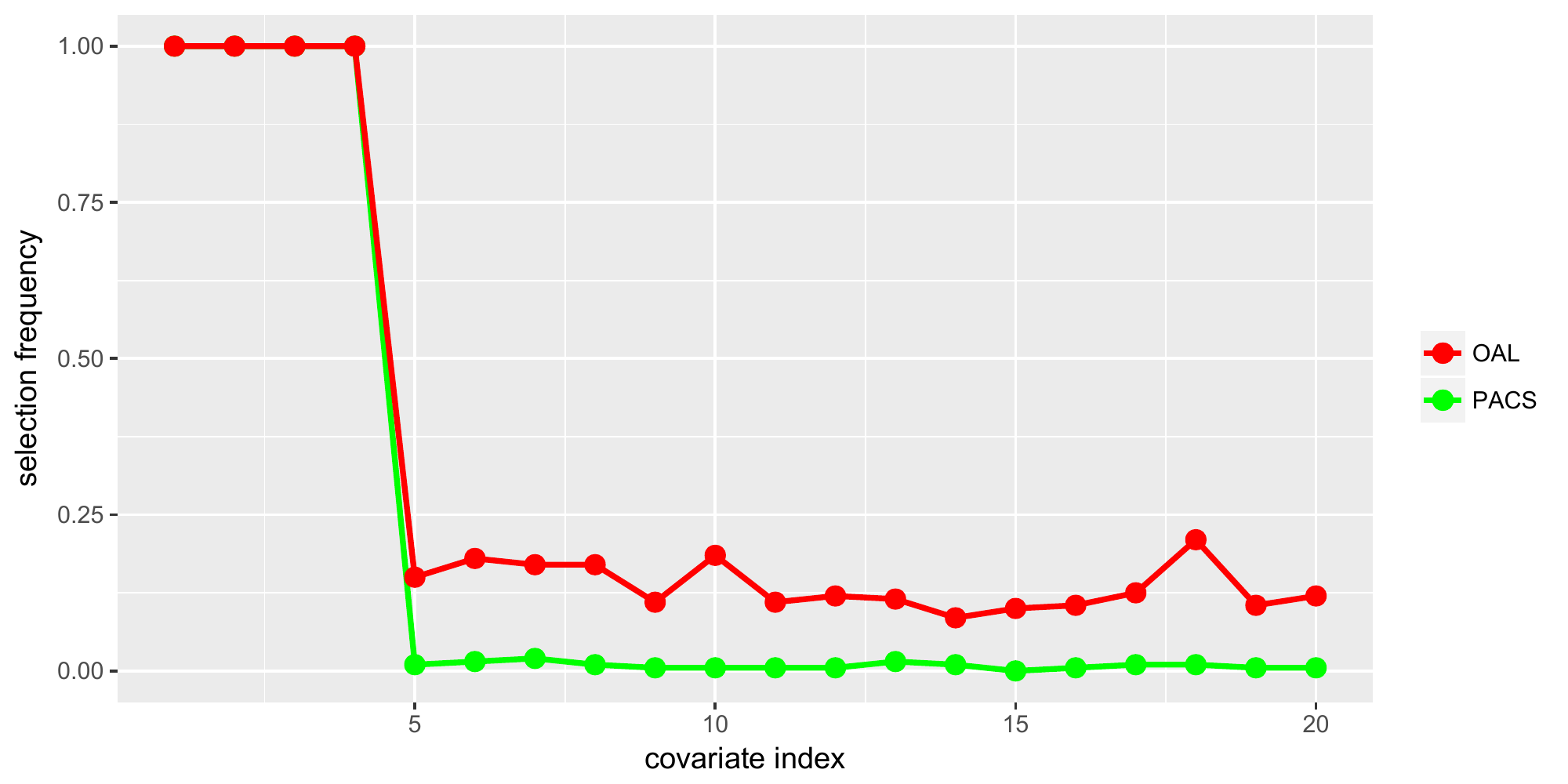}
		\includegraphics[width=22em]{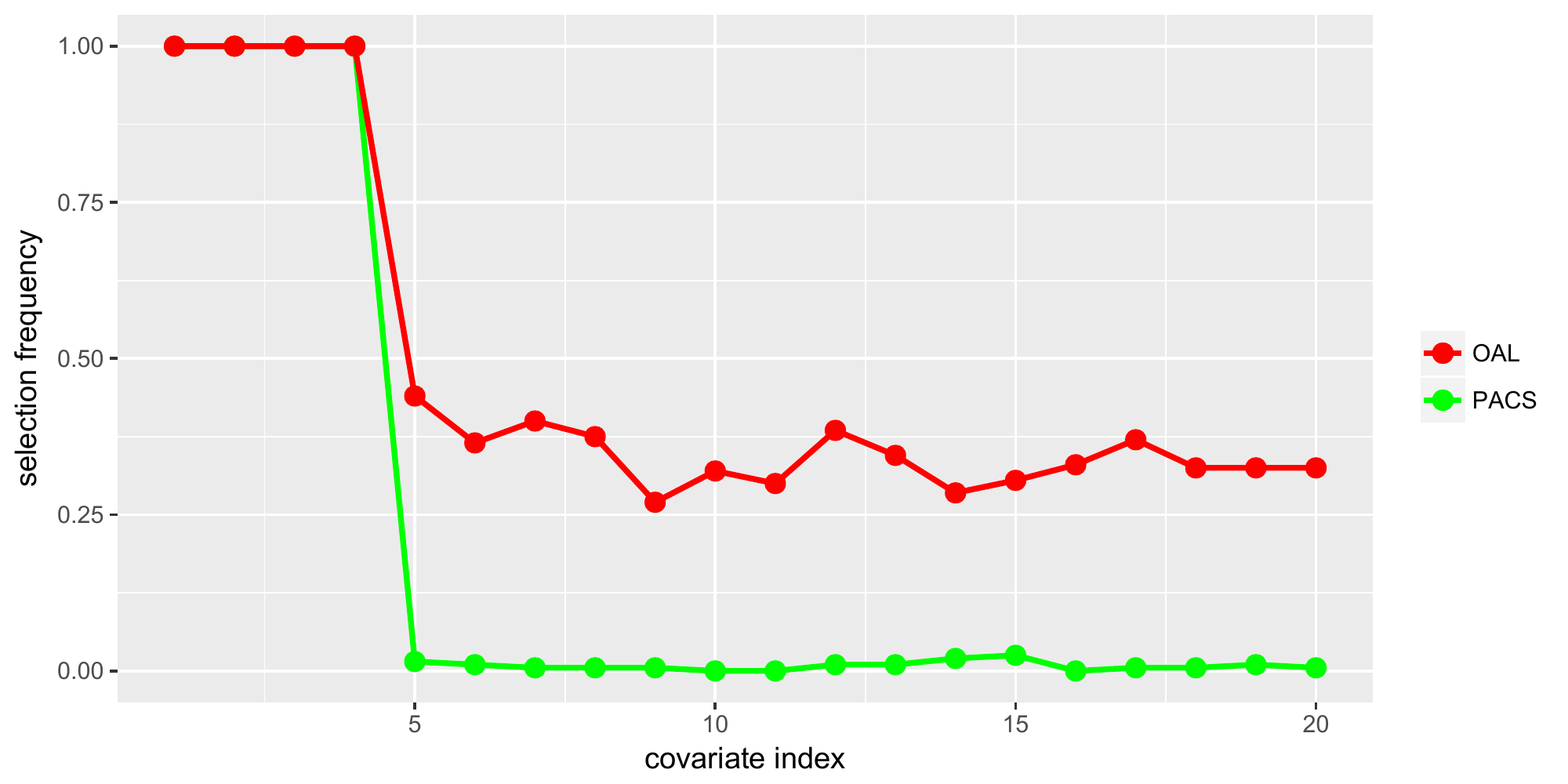}
	\end{center}
    \begin{center}
    	Figure 2. Compare PACS and OAL: Frequency of variable selected into the model in Scenario 1 (model misspecification), $\alpha=(0.4, 0.4, 0, 0, 1, 1, 1, 1, 0, \cdots, 0)^\top$ (weak confounders).
    \end{center}

    \vspace{15mm}
    
    \begin{center}
    	$\beta^T=(0.6, 0.6, 0.8, 0.8, 0, \cdots, 0)^\top$ \hspace{26mm} $\beta^T=(0.6, 0.6, 1.2, 1.2, 0, \cdots, 0)^\top$ \hspace{26mm} $\beta^T=(0.6, 0.6, 2.4, 2.4, 0, \cdots, 0)^\top$ \\
    	$\beta^C=(0.8, 0.8, 0.6, 0.6, 0, \cdots, 0)^\top$ \hspace{26mm} $\beta^C=(1.2, 1.2, 0.6, 0.6, 0, \cdots, 0)^\top$ \hspace{26mm} $\beta^C=(2.4, 2.4, 0.6, 0.6, 0, \cdots, 0)^\top$
    \end{center}
    \begin{center}
    	\includegraphics[width=22em]{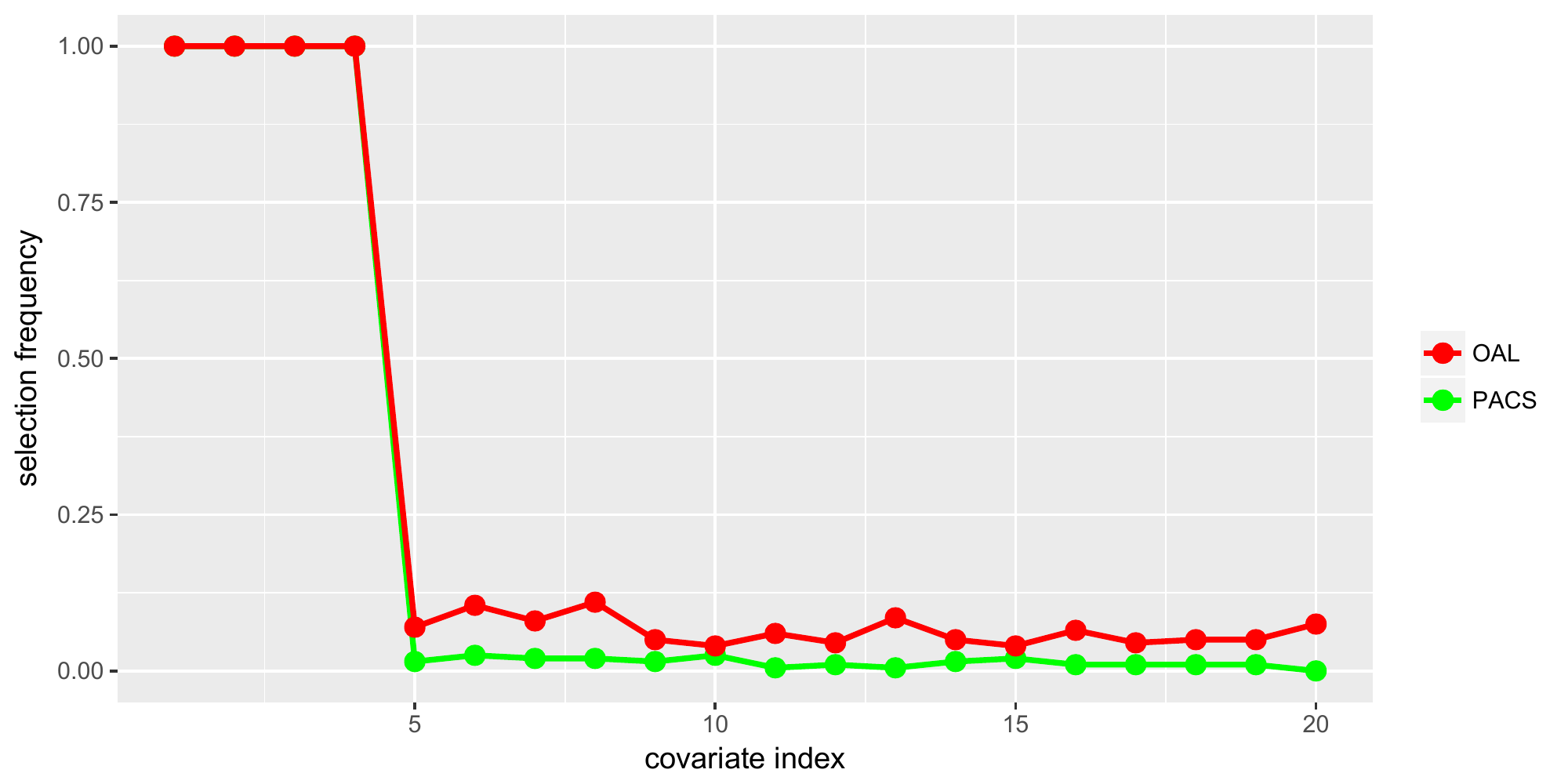}
    	\includegraphics[width=22em]{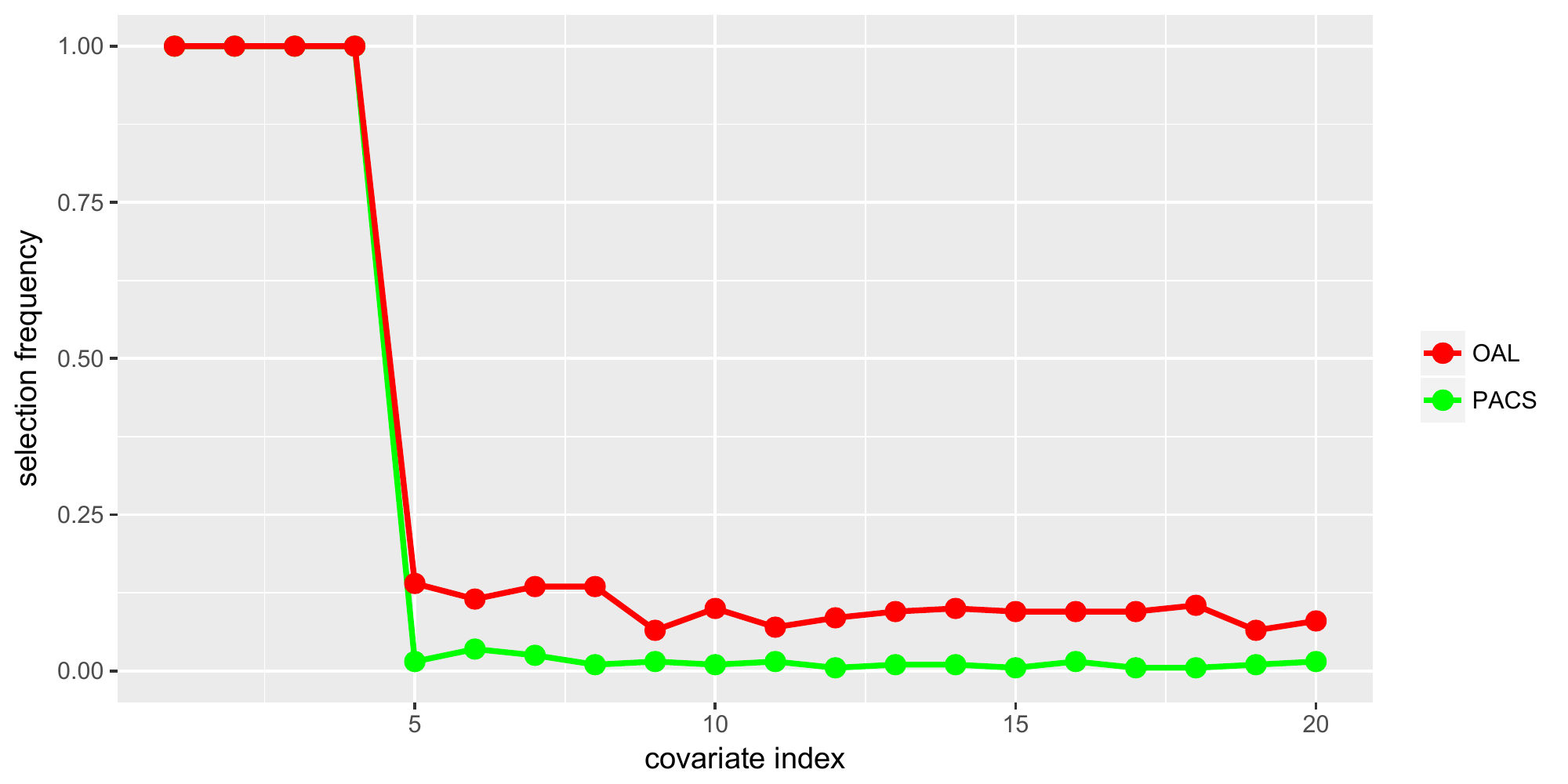}
    	\includegraphics[width=22em]{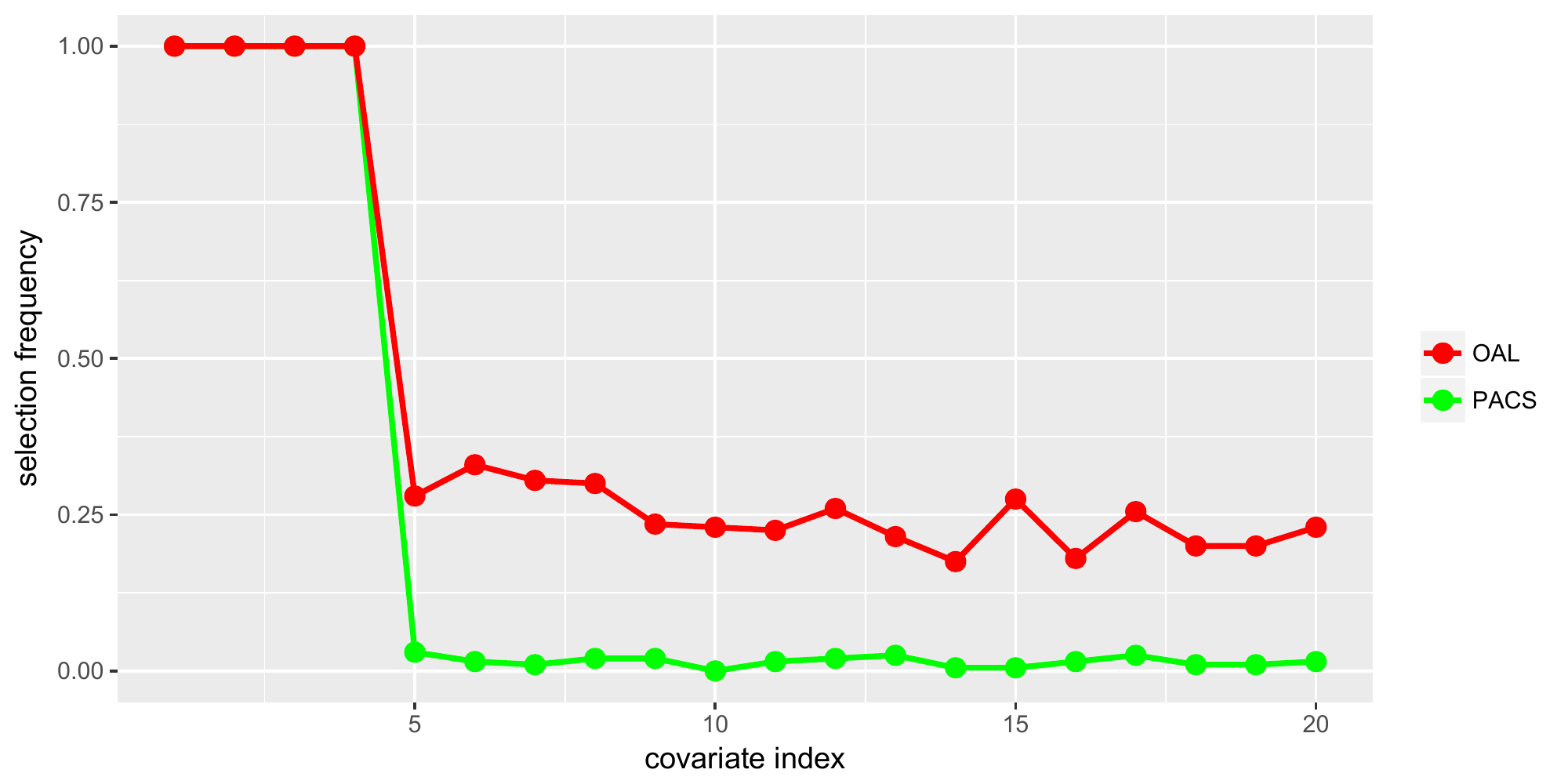}
    \end{center}
    \begin{center}
    	Figure 3. Compare PACS and OAL: Frequency of variable selected into the model in Scenario 1 (model misspecification), $\alpha=(1, 1, 0, 0, 1, 1, 1, 1, 0, \cdots, 0)^\top$ (strong confounders).
    \end{center}
\end{landscape}

\begin{landscape}
	\begin{center}
		$n=500$ \hspace{66mm} $n=500$ \hspace{63mm} $n=1000$ \\
		$p=100$ \hspace{66mm} $p=20$ \hspace{69mm} $p=20$
	\end{center}
	\begin{center}
		\includegraphics[width=22em]{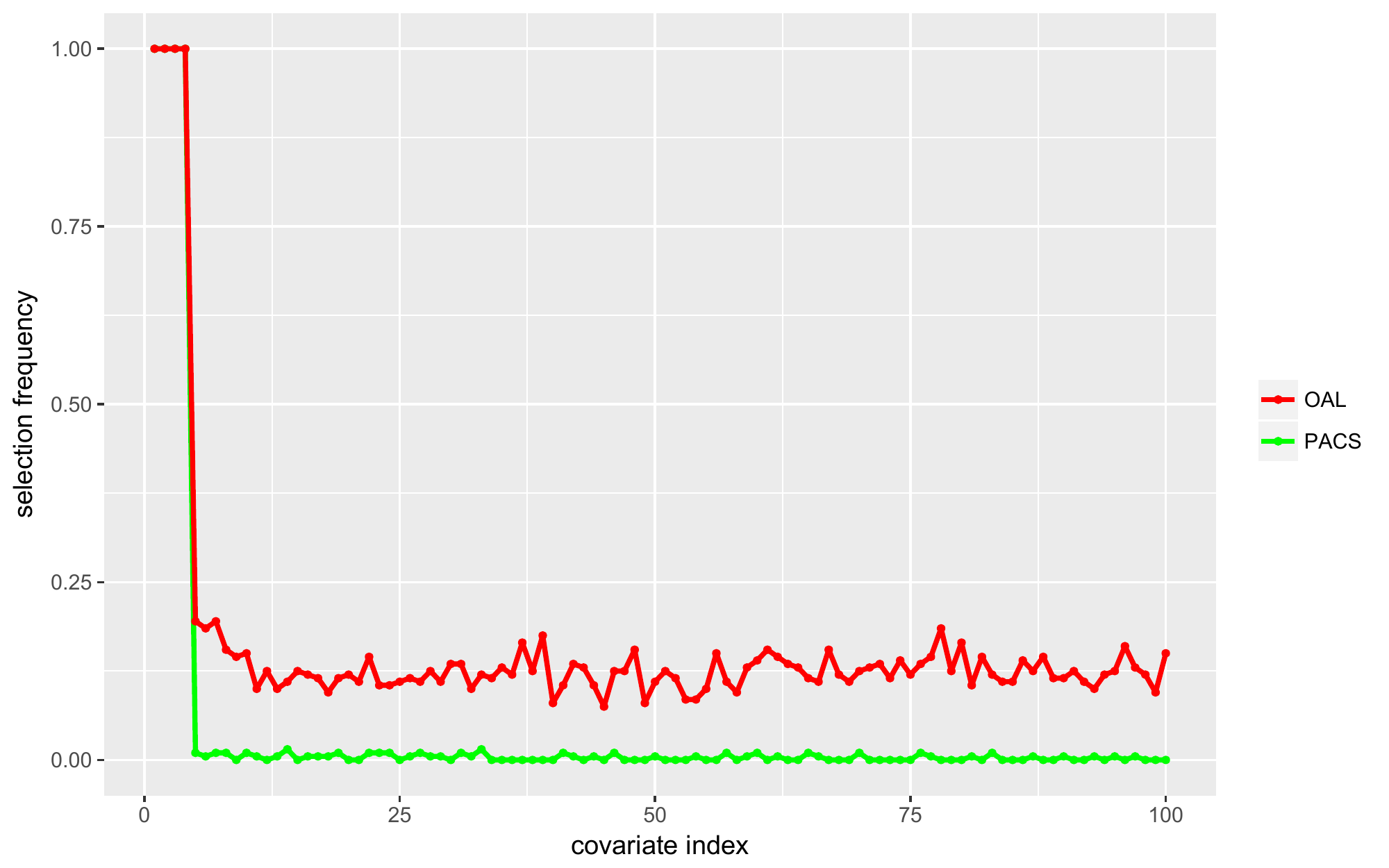}
		\includegraphics[width=22em]{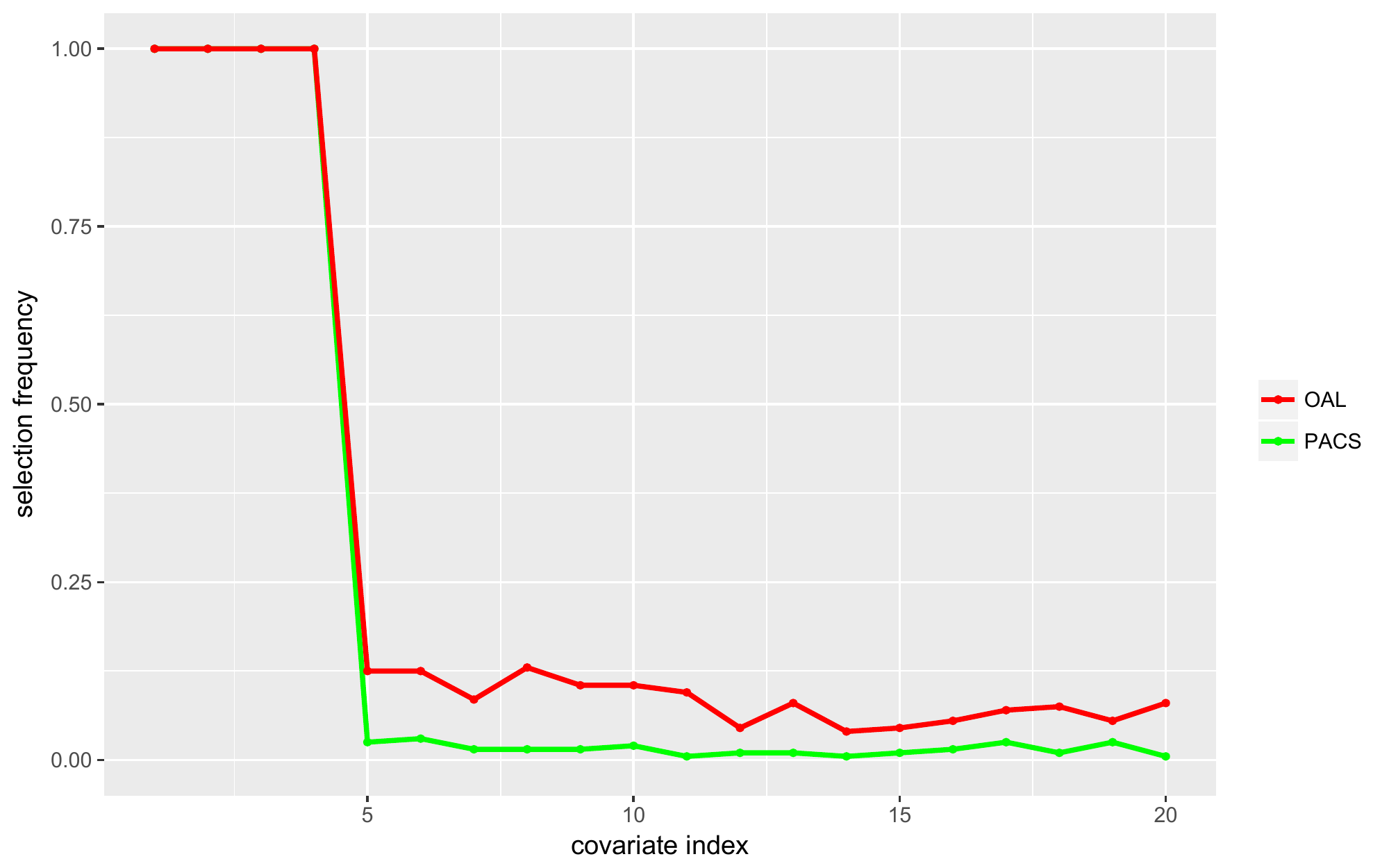}
		\includegraphics[width=22em]{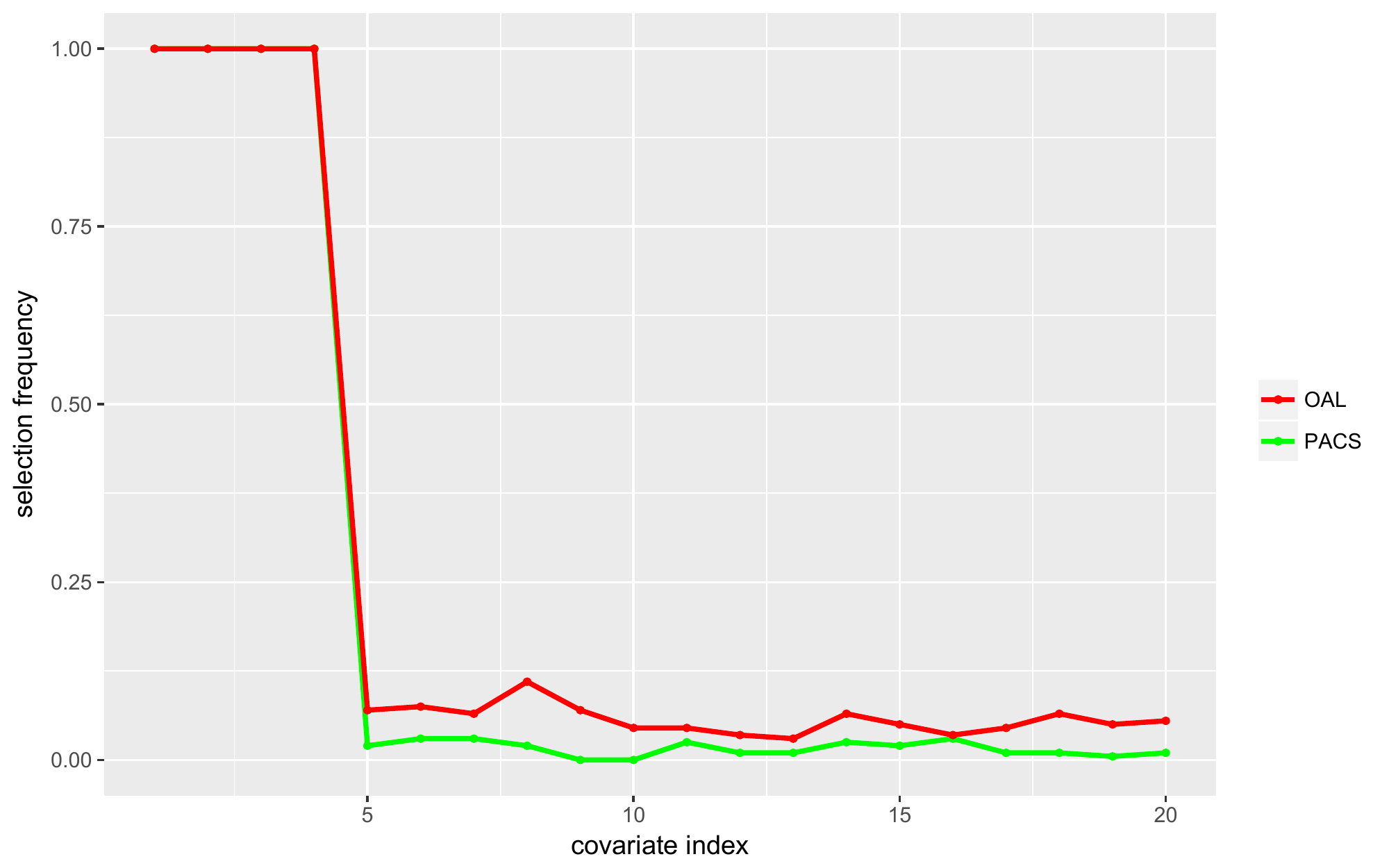}
	\end{center}
	\begin{center}
		Figure 4. Compare PACS and OAL: Frequency of variable selected into the model in Scenario 2 (linear outcome models), $\alpha=(0.4, 0.4, 0, 0, 1, 1, 1, 1, 0, \cdots, 0)^\top$ (weak confounders).
	\end{center}
	
	\vspace{10mm}
	
	\begin{center}
		$n=500$ \hspace{66mm} $n=500$ \hspace{63mm} $n=1000$ \\
		$p=100$ \hspace{66mm} $p=20$ \hspace{69mm} $p=20$
	\end{center}
	\begin{center}
		\includegraphics[width=22em]{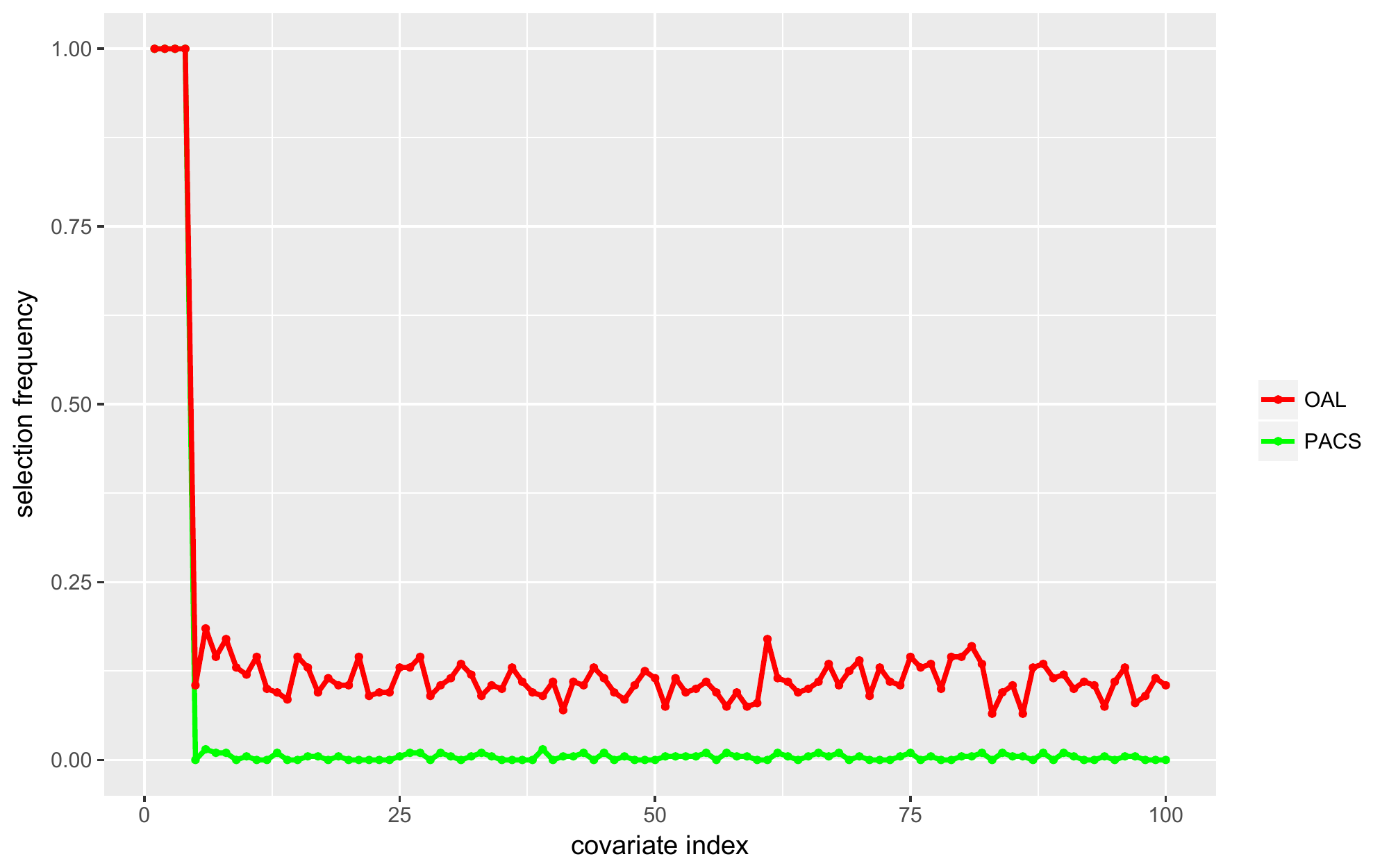}
		\includegraphics[width=22em]{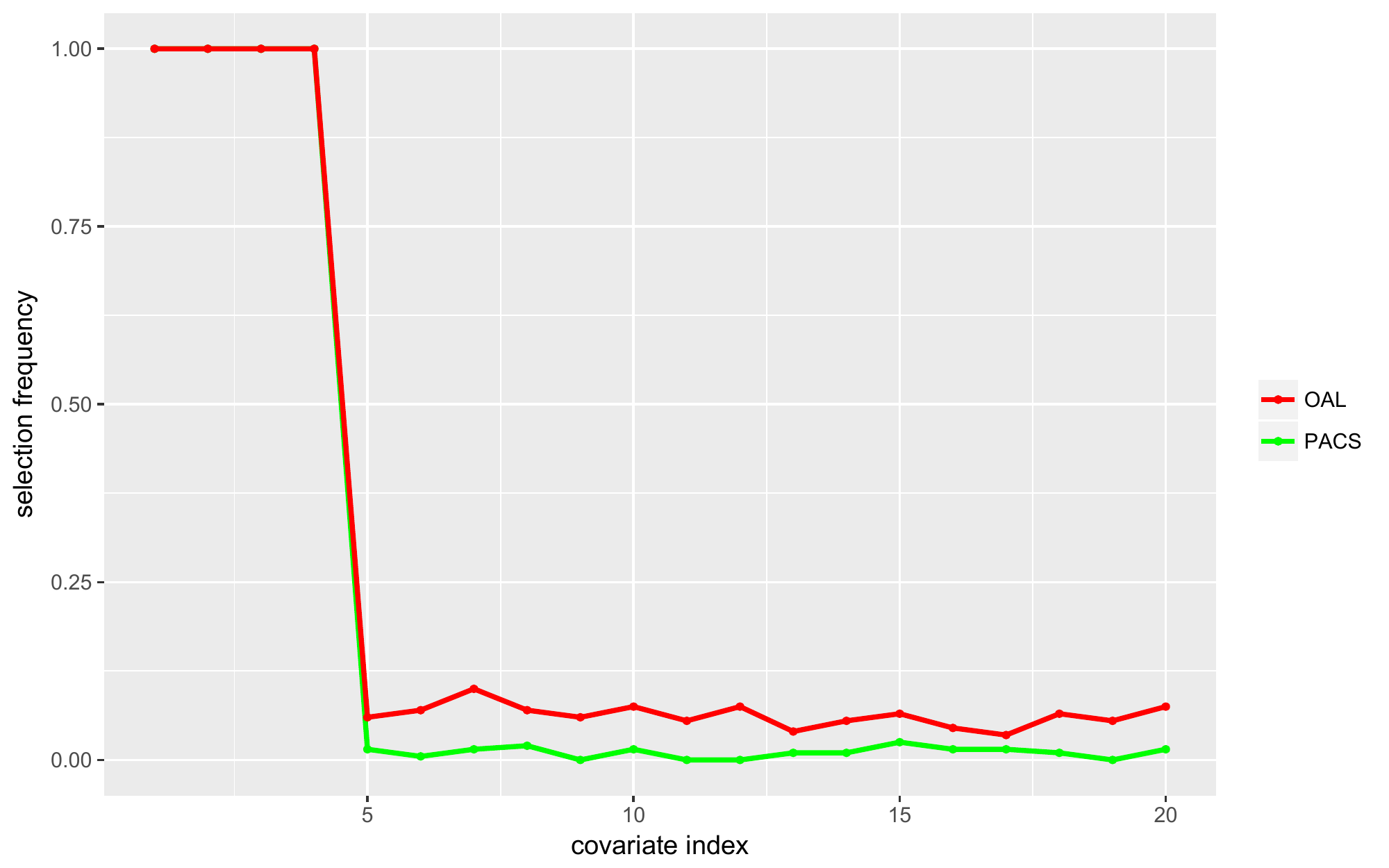}
		\includegraphics[width=22em]{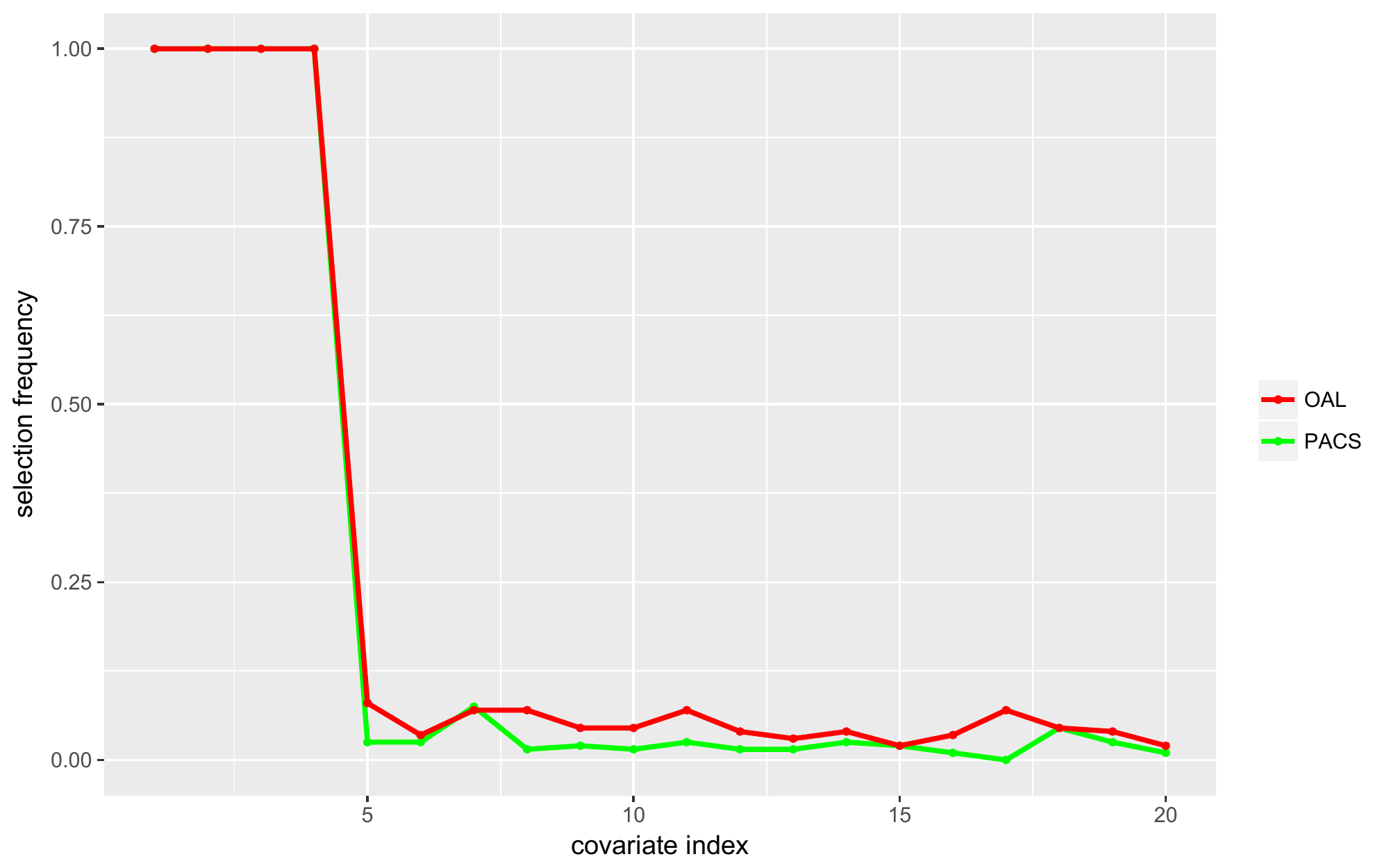}
	\end{center}
	\begin{center}
		Figure 5. Compare PACS and OAL: Frequency of variable selected into the model in Scenario 2 (linear outcome models), $\alpha=(1, 1, 0, 0, 1, 1, 1, 1, 0, \cdots, 0)^\top$ (strong confounders).
	\end{center}
\end{landscape}

\section{Discussion}
\noindent In this article, we propose a propensity score adapted covariate selection (PACS) procedure, which is robust to outcome model misspecification under the ``linear association" conditions. Due to its adequate use of information from both treatment group and control group, PACS is more efficient in excluding instrumental variables and spurious covariates compared to previous approaches. We also point out that the key part is covariate selection rather than propensity score estimation, hence a good covariate selection method should mainly focus on outcome model. Results of simulation studies are presented to support our theoretical analysis.

There are some future research directions towards which our work can be further extended and enhanced. First, the performance of PACS will be not as perfect as illustrated in Section 4, when the outcome model is seriously misspecified. For examples, if generalized linear models with non-linear link functions (exponential or trigonometric) for $Y^T$ and/or $Y^C$ are assumed, then PACS may exclude $\mathbf{X}_{\mathcal{A}}$ with frequency approximately $5\%$, but still has a lower rate of including $\mathbf{X}_{\mathcal{A}^c}$ than the OAL. Under such situation, we suggest that practitioners use an alternative strategy:
\begin{center}
	For $j=1, \cdots, p$, if $\hat{\beta}_{PACS, j}^T\neq 0$ or $\hat{\beta}_{PACS, j}^C \neq 0$, we select covariate $X_j$ into the propensity score model.
\end{center}
Second, it is possible that the potential outcomes $(Y^T, Y^C)$ do depend on the target covariates $\mathbf{X}_{\mathcal{A}}$ but the ``linear association" conditions are violated, although of little chance. In this case, the PACS will include nothing into the propensity score model, neither do other approaches based on the correct specification of linear outcome models. Therefore explorations of oracle procedures under weaker conditions should not stop here. Additionally, it is necessary to design a general criterion in detecting whether the ``linear association" conditions hold true in studies of real data. Finally, we use a parametric (usually logistic or probit) model to estimate the propensity scores in the first step of implementation of PACS. For the consideration of convergence, $p$ should not be too large compared to $n$. However, we often have to face a great (sometimes even diverging) amount of covariates, hence it is also of great importance to generalize our theory to deal with the case in which both $n$ and $p$ go to infinity with a proper manner.

\section{Appendix}
\noindent \begin{proof}[Proof of Theorem 1] We only need to show that under Condition 1, $\hat{\beta}_{PACS}^T$ possess oracle properties, the other half will follow similarly. We first examine the least false parameter $\beta^{T*}$, note that $\hat{p}(\mathbf{X})$ is a consistent estimator of $p(\mathbf{X})$, letting $n \rightarrow \infty$ in $(1)$, we obtain
\begin{align*}
	\left( \beta^{T*}, \eta^{T*} \right) = & \argmin_{\beta, \eta} \mathrm{E} \left[ \frac{D}{p(\mathbf{X})} (Y-\eta-\mathbf{X}^\top \beta)^2 \right] \\
	= & \argmin_{\beta, \eta} \mathrm{E} \left[ \mathrm{E} \left[ \frac{D}{p(\mathbf{X})} (Y^T-\eta-\mathbf{X}^\top \beta)^2 \Big\vert \mathbf{X} \right] \right] \\
	= & \argmin_{\beta, \eta} \mathrm{E} \left[ (Y^T-\eta-\mathbf{X}^\top \beta)^2 \right].
\end{align*}
Therefore $\beta^{T*} = \mathrm{Cov}(\mathbf{X}, \mathbf{X})^{-1} \mathrm{Cov}(\mathbf{X}, Y^T)$. Now since $\mathbf{X}_{\mathcal{A}} \perp \mathbf{X}_{\mathcal{A}^c}$, as assumed, we have
\begin{equation}
	\beta^{T*}_{\mathcal{A}} = \mathrm{Cov}(\mathbf{X}_{\mathcal{A}}, \mathbf{X}_{\mathcal{A}})^{-1} \mathrm{Cov}(\mathbf{X}_{\mathcal{A}}, Y^T), \ \beta^{T*}_{\mathcal{A}^c} = \mathrm{Cov}(\mathbf{X}_{\mathcal{A}^c}, \mathbf{X}_{\mathcal{A}^c})^{-1} \mathrm{Cov}(\mathbf{X}_{\mathcal{A}^c}, Y^T),        \tag*{}
\end{equation}
where $\beta^{T*}_{\mathcal{A}}$ denotes the first $p_0$ components of $\beta^{T*}$, $\beta^{T*}_{\mathcal{A}^c}$ denotes the latter $p-p_0$ components of $\beta^{T*}$. 

According to Condition 1, $\forall j \in \mathcal{A}$, $\beta^{T*}_j \neq 0$. Since $\mathbf{X}_{\mathcal{A}^c} \perp Y^T$ (Exclusion Restriction), $\forall j \in \mathcal{A}^c$, $\beta^{T*}_j = 0$. Hence $\beta^{T*}$ is sign-consistent. Now we begin proving the oracle properties, we first show the asymptotic normality of $\hat{\beta}_{PACS}^T$. Let $\hat{u} = \sqrt{n}(\hat{\beta}_{PACS}^T - \beta^{T*})$, $\hat{v}=\sqrt{n}(\hat{\eta}_{PACS}^T - \eta^{T*})$, denote $z = (u^\top, v)^\top$, and define
\begin{equation}
	\Phi_n(u, v) = \sum_{i \in T} \frac{1}{\hat{p}(\mathbf{X}_i)} \left(Y_i - \left(\eta^{T*}+\frac{v}{\sqrt{n}}\right) - \mathbf{X}_i^\top \left(\beta^{T*} + \frac{u}{\sqrt{n}}\right)\right)^2 + \lambda_n \sum_{j=1}^{p} \hat{\omega}_j^T \left\vert \beta_j^{T*} + \frac{u_j}{\sqrt{n}} \right\vert,   \tag*{}
\end{equation}
then we have
\begin{align*}
	R_n(u, v) = \Phi_n(u, v) - \Phi_n(0, 0) = & z^\top \left( \frac{1}{n} \sum_{i=1}^{n} \frac{D_i}{\hat{p}(\mathbf{X}_i)} \begin{pmatrix}
	\mathbf{X}_i \\ 1 \end{pmatrix} \left(\mathbf{X}_i^\top, 1 \right) \right)z \\
	& - 2 \frac{1}{\sqrt{n}} \sum_{i=1}^{n} \frac{D_i}{\hat{p}(\mathbf{X}_i)} \left(Y_i - \eta^{T*} - \mathbf{X}_i^\top \beta^{T*} \right) \left(\mathbf{X}_i^\top, 1 \right) z \\
	& + \frac{\lambda_n}{\sqrt{n}} \sum_{j=1}^{p} \hat{\omega}_j^T \sqrt{n} \left(\left\vert \beta_j^{T*} + \frac{u_j}{\sqrt{n}} \right\vert - \left\vert \beta_j^{T*}\right\vert\right).
\end{align*}
Note that since $\hat{p}(\mathbf{X}_i)$ is consistent, there exists a constant matrix $\mathbf{T}$, such that
\begin{equation}
	\frac{1}{n} \sum_{i=1}^{n} \frac{D_i}{\hat{p}(\mathbf{X}_i)} \begin{pmatrix}
	\mathbf{X}_i \\ 1 \end{pmatrix} \left(\mathbf{X}_i^\top, 1 \right) \rightarrow \mathbf{T}, \ n \rightarrow \infty.  \tag*{}
\end{equation}
Since $p(\mathbf{X})$ is from a parametric model with parameter $\alpha$ (as discussed in Section 3.1), we have
\begin{equation}
	\hat{p}(\mathbf{X})-p(\mathbf{X}) = (\hat{\alpha}-\alpha)^\top \frac{\partial p}{\partial \alpha} (\mathbf{X}), \tag*{} 
\end{equation}
and further
\begin{align*}
	& \frac{1}{\sqrt{n}} \sum_{i=1}^{n} \frac{D_i}{\hat{p}(\mathbf{X}_i)} \left(Y_i - \eta^{T*} - \mathbf{X}_i^\top \beta^{T*} \right) \left(\mathbf{X}_i^\top, 1 \right) = \frac{1}{\sqrt{n}} \sum_{i=1}^{n} \frac{D_i}{p(\mathbf{X}_i)} \left(Y_i - \eta^{T*} - \mathbf{X}_i^\top \beta^{T*} \right) \left(\mathbf{X}_i^\top, 1 \right)  \\
	& - \sqrt{n} (\hat{\alpha}-\alpha)^\top \frac{1}{n} \sum_{i=1}^{n} \frac{D_i}{p(\mathbf{X}_i)^2} \frac{\partial p}{\partial \alpha} (\mathbf{X}_i) \left(Y_i - \eta^{T*} - \mathbf{X}_i^\top \beta^{T*} \right) \left(\mathbf{X}_i^\top, 1 \right) + O_p\left(\frac{1}{\sqrt{n}}\right)  \\
	= & \frac{1}{\sqrt{n}} \sum_{i=1}^{n} \frac{D_i}{p(\mathbf{X}_i)} \left(Y_i - \eta^{T*} - \mathbf{X}_i^\top \beta^{T*} \right) \left(\mathbf{X}_i^\top, 1 \right)  \\
    & - \sqrt{n} (\hat{\alpha}-\alpha)^\top \mathrm{E} \left[ \frac{D}{p(\mathbf{X})^2} \frac{\partial p}{\partial \alpha} (\mathbf{X}) \left(Y - \eta^{T*} - \mathbf{X}^\top \beta^{T*} \right) \left(\mathbf{X}^\top, 1 \right) \right] + O_p\left(\frac{1}{\sqrt{n}}\right) .
\end{align*}
Now since $\sqrt{n}(\hat{\alpha}-\alpha)$ converges to some normal distribution as $n \rightarrow \infty$, using Slutsky's theorem, we know that there exists a non-negative symmetric matrix $\mathbf{\Sigma}$,
\begin{equation}
	\frac{1}{\sqrt{n}} \sum_{i=1}^{n} \frac{D_i}{\hat{p}(\mathbf{X}_i)} \left(Y_i - \eta^{T*} - \mathbf{X}_i^\top \beta^{T*} \right) \left(\mathbf{X}_i^\top, 1 \right) \rightarrow_{d.} \mathcal{N}(0, \mathbf{\Sigma}), \ n \rightarrow \infty.  \tag*{}
\end{equation}
Then we turn to the third term, if $\beta_j^{T*} \neq 0$, then $\hat{\omega}_j^T \rightarrow \left\vert \beta_j^{T*} \right\vert^{-\gamma}$ in probability, hence
\begin{equation}
	\frac{\lambda_n}{\sqrt{n}} \hat{\omega}_j^T \sqrt{n} \left(\left\vert \beta_j^{T*} + \frac{u_j}{\sqrt{n}} \right\vert - \left\vert \beta_j^{T*}\right\vert\right) \rightarrow_{p} 0, \ n \rightarrow \infty.  \tag*{}
\end{equation}
If $\beta_j^{T*} = 0$, then $1 = O_p(\hat{\omega}_j^T n^{-\gamma/2})$, therefore if $u_j \neq 0$, then we have
\begin{equation}
	\frac{\lambda_n}{\sqrt{n}} \hat{\omega}_j^T \left\vert u_j \right\vert = (\lambda_n n^{(\gamma-1)/2}) (\hat{\omega}_j^T n^{-\gamma/2}) \left\vert u_j \right\vert \rightarrow \infty, \ n \rightarrow \infty.  \tag*{}
\end{equation}
For any bounded $z = (u^\top, v)^\top$, if $u_{\mathcal{A}^c} = 0$, then $R_n(u, v) \rightarrow z^\top \mathbf{T} z - 2 \mathcal{N}(0, \mathbf{\Sigma}) z$ in distribution, otherwise $R_n(u, v) \rightarrow \infty$. Now since $(\hat{u}^\top, \hat{v})^\top$ minimizes $R_n(u, v)$, whereas $\begin{pmatrix}
\mathbf{T}_{\mathcal{A}}^{-1} \mathcal{N}(0, \mathbf{\Sigma}_{\mathcal{A}}) \\ 0 \\ v^*
\end{pmatrix}$ minimizes $z^\top \mathbf{T} z - 2 \mathcal{N}(0, \mathbf{\Sigma}) z$, following the epi-convergence argument of Geyer (1994)\cite{geyer1994asymptotics} and Knight and Fu (2000)\cite{knight2000asymptotics}, we can prove the asymptotic normality of $\hat{u}$ and $\hat{v}$.

Then we turn to prove consistency of variable selection. Based on asymptotic normality of $\hat{\beta}_{PACS}^T$ and sign-consistency of $\beta^{T*}$, $\lim_{n \rightarrow \infty} \mathrm{P} (\hat{\beta}_{PACS, j}^T \neq 0, \ \forall j \in \mathcal{A} = \mathcal{U} \cup \mathcal{C}) = 1$ is automatically deduced. We only need to show that $\mathrm{P} (\hat{\beta}_{PACS, j}^T=0) \rightarrow 1$ as $n \rightarrow \infty$, $\forall j \in \mathcal{A}^c$.

If $\hat{\beta}_{PACS, j}^T \neq 0$, then the KKT optimality conditions tell us
\begin{equation}
	\left\vert \frac{1}{\sqrt{n}} \sum_{i = 1}^{n} \frac{D_i}{\hat{p}(\mathbf{X}_i)} 2 X_{ij} \left(Y_i - \hat{\eta}_{PACS}^T - \mathbf{X}_i^\top\hat{\beta}_{PACS}^T \right)\right\vert = \frac{\lambda_n}{\sqrt{n}} \hat{\omega}_j^T = \frac{\lambda_n n^{(\gamma-1)/2}}{\left(\sqrt{n} \left\vert \tilde{\beta}_j^T \right\vert\right)^{\gamma}}.    \tag*{}
\end{equation}
Since $(\hat{\beta}_{PACS}^T, \hat{\eta}_{PACS}^T)$ is asymptotically normal, $\hat{p}(\mathbf{X}_i)$ is $\sqrt{n}$-consistent, we know that
\begin{equation}
	\xi_n = \frac{1}{\sqrt{n}} \sum_{i = 1}^{n} \frac{D_i}{\hat{p}(\mathbf{X}_i)} 2 X_{ij} \left(Y_i - \hat{\eta}_{PACS}^T - \mathbf{X}_i^\top\hat{\beta}_{PACS}^T \right)   \tag*{}
\end{equation}
converges to some normal distribution as $n \rightarrow \infty$. But $\lambda_n n^{(\gamma-1)/2} \Big/ \left(\sqrt{n} \left\vert \tilde{\beta}_j^T \right\vert\right)^{\gamma} \rightarrow \infty$ in probability, hence we have
\begin{equation}
    \mathrm{P} \left(\hat{\beta}_{PACS, j}^T \neq 0\right) \leq \mathrm{P}\left(\xi_n = \frac{\lambda_n n^{(\gamma-1)/2}}{\left(\sqrt{n} \left\vert \tilde{\beta}_j^T \right\vert\right)^{\gamma}}\right) \rightarrow 0, \ n \rightarrow \infty.   \tag*{}
\end{equation}
We complete the proof of covariate selection consistency. The other parts can be shown in the same way.
\end{proof}

\bibliographystyle{plain}
\bibliography{Mybib}

\end{document}